\documentclass[twocolumn]{aastex631}

\def\deg{${}^\circ$}

\def\sec{${}^{\prime\prime}$}

\shorttitle{Star Forming Regions in the Galactic Plane with Rubin LSST}
\shortauthors{Prisinzano et al.}

\graphicspath{{./}{figures/}}

\begin{document}

\title{Rubin LSST  observing strategies to maximize volume and uniformity coverage of  Star Forming Regions in the Galactic 
Plane\footnote{\bf {to be considered for the ApJS  Rubin Cadence Focus Issues}}}

\author[0000-0002-8893-2210]{Loredana Prisinzano}
\affiliation{INAF-Osservatorio Astronomico di Palermo, Piazza del Parlamento, 1, 90129, Palermo, Italy}

\author[0000-0001-9297-7748]{Rosaria Bonito}
\affiliation{INAF-Osservatorio Astronomico di Palermo, Piazza del Parlamento, 1, 90129, Palermo, Italy}


\author[0000-0002-7503-5078]{Alessandro Mazzi}
\affiliation{Dipartimento di Fisica e Astronomia Galileo Galilei, Universit\`a di Padova, Vicolo dell’Osservatorio 3, I-35122 Padova, Italy}

\author[0000-0002-7065-3061]{Francesco Damiani}
\affiliation{INAF-Osservatorio Astronomico di Palermo, Piazza del Parlamento, 1, 90129, Palermo, Italy}

\author[0000-0003-4596-2628]{Sabina Ustamujic}
\affiliation{INAF-Osservatorio Astronomico di Palermo, Piazza del Parlamento, 1, 90129, Palermo, Italy}

\author[0000-0003-2874-6464]{Peter Yoachim}
\affiliation{ Department of Astronomy, University of Washington, Box 351580, Seattle, WA 98195, USA}

\author[0000-0001-6279-0552]{Rachel Street}
\affiliation{Las Cumbres Observatory, 6740 Cortona Dr., Suite 102, Goleta, CA 93117, USA}

\author[0000-0002-3010-2310]{Mario Giuseppe Guarcello}
\affiliation{INAF-Osservatorio Astronomico di Palermo, Piazza del Parlamento, 1, 90129, Palermo, Italy}

\author[0000-0002-4115-0318]{Laura Venuti}
\affiliation{SETI Institute, 339 Bernardo Avenue, Suite 200, Mountain View, CA 94043, USA}

\author[0000-0002-2577-8885]{William Clarkson}
\affiliation{Department of Natural Sciences, University of Michigan-Dearborn, 4901 Evergreen Rd, Dearborn, MI 48128, USA}

\author[0000-0001-5916-0031]{Lynne Jones}
\affiliation{Department of Astronomy, University of Washington, Box 351580, Seattle, WA 98195, USA}

\author[0000-0002-6301-3269]{Leo Girardi}
\affiliation{INAF – Osservatorio Astronomico di Padova, Vicolo dell’Osservatorio 5, I-35122 Padova, Italy}



\begin{abstract}
A complete map of the youngest stellar populations of the Milky Way 
in the era of all-sky surveys,
is one of the  most challenging  goals in modern astrophysics.
The characterization of the youngest stellar component is crucial not only for a global overview of the Milky Way structure,  of the Galactic thin disk, and its spiral arms, but also for local studies. In fact,
the identification of the star forming regions (SFRs) and the comparison with the environment in which they form are also fundamental to put them in the context of the surrounding giant molecular clouds and to understand
still unknown  physical mechanisms related to the star and planet formation
processes. In 10 yrs of observations, Vera C. Rubin Legacy Survey of Space and Time (Rubin-LSST) will achieve an exquisite photometric  depth that will
allow us to significantly extend the volume within which we will be able to discover
new SFRs and to enlarge the domain of a detailed knowledge of our own Galaxy. 
We describe here a metrics that estimates the total number of young stars with ages $t<10$\,Myr and masses $>$0.3\,M$_\odot$ that will be detected with the Rubin LSST observations in the gri bands at a 5\,$\sigma$ magnitude significance. We examine  the results of our metrics adopting the most recent simulated Rubin-LSST survey strategies in order to evaluate the impact that different  observing strategies 
might have on our science case.

\end{abstract}



\section{Introduction} \label{sec:intro}
In the era of  all-sky astrophysical surveys, mapping the youngest stellar populations of the Milky Way 
in the optical bands is one of the 
 main core science goals.
Up to now, a full overall understanding of the Galactic components has been hampered by observational limits and, paradoxically, we have a better view of the morphological structure of external galaxies than of the Milky Way.
In particular, it is crucial to trace the poorly known   Galactic thin disk,
and its spiral arms, where most of the youngest stellar populations
are expected to be found.

The youngest galactic  stellar component is mainly found in star forming regions (SFRs), namely stellar clusters and over-dense structures originating from the collapse of the molecular clouds, the coldest and densest part of the interstellar medium \citep{macl04}. 
It is by now relatively easy to identify in the 
near-, mid-, far-infrared (IR), and radio wavelengths the young stellar objects (YSOs) within SFRs, during the first phases of the star formation process. In fact, at these wavelengths, due to the presence of the optically thick infalling  envelope or circumstellar disk around the central star, they show, at these wavelengths, an excess with respect to the typical photospheric emission. During the  subsequent phases, they start to emit also in the optical bands,
while the circumstellar disk is still optically thick \citep{bouv07}. YSOs are no longer easily identifiable  in the IR or at radio wavelengths, when the final dispersal of the disk material occurs and 
non-accreting transition disks  form \citep{erco21}. At these latter stages,
a complete census of the YSOs can be achieved only using  deep optical
observations. Such census
is therefore  crucial for a thorough comprehension  of the large scale three-dimensional structure of the young Galactic stellar component.

The identification of SFRs and the comparison with the environment in which they form are also fundamental to put them in the context of the surrounding giant molecular clouds and to answer 
 still controversial open questions about the star formation (SF) process,
 such as, i) did the SFRs we observe originate from monolithic single SF bursts or from multiple spatial and temporal scales, 
i.e. in a hierarchical mode \citep[e.g.][]{kamp08}? ii) When can the feedback from massive stars affect the proto-planetary disk evolution \citep{tana18} or 
trigger subsequent SF events,
as thought, for example, for the Supernova explosion in the $\lambda$\,Ori region \citep{koun20}? 
iii) Is the overall spatial 
distribution of SFRs correlated to an overall large scale structure, such as the Gould Belt or the damped wave, very recently suggested by \citet{alve20}?

The Vera C. Rubin Observatory \citep{ivez19}
will conduct a Legacy Survey of Space and Time (LSST) 
with a very impressive combination of flux sensitivity, area, and temporal sampling rate \citep{bian22}. 

A flexible scheduling system is designed to maximize the scientific return under a set of observing constraints. Vera C. Rubin will perform in the 10 years a set of surveys. These include the Wide-Fast-Deep (WFD) that is the “main” LSST survey, primarily focused on low-extinction regions of the sky for extra-galactic science, the Deep Drilling Fields (DDFs), that are a set of few individual fields that will receive high cadence 
and additional “minisurveys”, that cover specific sky regions such as the ecliptic plane, Galactic plane (GP), and the Large and Small Magellanic Clouds
with specific observing parameters \citep{bian22}.

The implementation of the Rubin LSST observing strategy must meet
the basic LSST Science Requirements\footnote{see \url{https://docushare.lsst.org/docushare/dsweb/Get/LPM-17}}
but  a significant flexibility in the detailed cadence of observations
is left to ensure the best optimisation of the survey.
 The scientific community has been involved to give 
feedback on specific science cases on 
the distribution of visits within a year, 
the distribution of images between filters, and the definition of a “visit” as single  or multiple exposures. 
To this aim, simulated realisations of the 10-yr sequence of observations acquired by LSST, for different sets of basis functions that define the 
different survey strategies, indicated as \texttt{OpSim}, are made available to the community. Such simulated observations can be analysed through 
an open-access software package, the metrics analysis framework \citep[MAF,][]{jone14},
where specific metrics can be calculated to quantify the suitability of a given strategy for reaching a particular science objective.

In the original baseline, i.e. the reference benchmark survey strategy, the mini survey of the GP was  covered by a number of visits a factor $\sim$5 smaller than those planned for WFD.
The main reason of this choice is to avoid  highly extincted regions with E(B-V)$>$0.2 since this amount of dust extinction is problematic for extragalactic science, as motivated in \citet{jone21}.
Nevertheless, mapping the Milky Way is one of the
four  broad science pillars that LSST will address \citep{bian22}.

We aim to show here that alternative implementations in which  the number of visits planned for the WFD is extended also to the Galactic thin disk ($|b|\lesssim $ 5-10\deg),
will increase significantly the impact of Rubin LSST on the understanding
of the Galactic large-scale structures and in particular of the large molecular clouds in which stars form. 
The stellar components of the SFRs,
and, in particular, the most populated
low mass components down to the M-dwarf regime with age t$\lesssim$10 Myrs,
that will be discovered with Rubin LSST observations,
are one of the most relevant output of the star formation
process and assessing their properties (spatial distribution in the Galaxy, masses, ages and age spread) is crucial to trace the global structure
of the youngest populations as well as to pinpoint 
several under-debate disputes such as the universality of the Initial Mass Function (IMF), the duration of the evolutionary phases involved in 
the star formation process and the dynamical evolution 
\citep{chab03,tass04,ball07,park14}.

Even though the thin disk of the Galaxy is 
strongly affected by high dust extinction and crowding issues,
it is the only Galactic component where most of such structures are found and where young stars are exclusively
found. 

A big step forward has been made in the  comprehension of the Galactic components thanks to the Gaia data
that allowed us to achieve a  clearer  and homogeneous overview of the
Galactic structures, including the youngest ones, but at distances from the Sun
limited to  2-3\,kpc 
\citep{zari18,koun19,koun20,kerr21,pris22}.
The exquisite science return from the very deep Rubin LSST static photometry, derived from co-added images, is the only opportunity to push such large-scale studies of young stars to otherwise unreachable and unexplored regions. 

With respect to the old open clusters, the study of the SFRs is favored by the intrinsic astrophysical properties of the low mass YSOs, representing the bulk
of these populations.
In fact, during the pre-main sequence (PMS) phase, M-type young stars are more luminous in bolometric light than the stars of the same mass in the main sequence (MS) phase. For example, for a given extinction, a 1\,Myr  (10 Myrs) old 0.3\,M$_\odot$ star  is 3.4\,mag (1.6\,mag) brighter than
a 100\,Myr star of the same mass  \citep{togn18,pris18a}.  
Therefore, if on one hand the extinction limits our capability to detect such low mass stars, on the other hand, the higher intrinsic luminosity of
young stars (t$<$10\,Myr) compensates for this effect, allowing a significant increase in
the volume of PMS stars detectable with respect to old MS stars. 
Such property has been exploited in \citet{dami18,pris18,venu19}
where a purely photometric approach using {\it gri} magnitudes combined with the most deep near-IR magnitudes available in the literature,  
has been adopted to statistically identify YSOs at distances where M-type MS stars are no longer
detectable, because they are intrinsically less luminous than the
analogous M-type PMS stars.

 We note, however, that also studies of other relevant structures, such as for example,  
stellar clusters, as tracers of the Galactic chemical evolution \citep{pris18}, as well as transient phenomena (see Street et al. 2021 Cadence Note) would benefit from the
observed strategy discussed in this paper.


The goal of this paper is to evaluate the impact on the YSO science  of different Rubin LSST observing strategies and, in particular, of the different WFD-like cadences extended to the GP, as an improvement with respect to the baseline observing strategy.

\section{metrics definition}
The metrics we present in this work is defined as the 
number of   detectable YSOs with masses down to 0.3 M$_\odot$ and ages t$\leq$10\,Myr, distributed in the GP and, in particular, in the thin disk of the Milky Way.

The number of  YSOs  estimated in this work is
not based on empirical results but on 
the  theoretical description 
of  the star number density $\rho$ (number of stars per unit volume) 
 in the Galactic thin disk, integrated within the volume accessible with Rubin LSST observations.  In particular,
the total number of young stars observable within a given solid angle $\Omega$  around a direction (e.g. a Healpix) defined by the Galactic coordinates $(l,b)$ will be obtained by integrating such density along cones, using volume elements
\begin{equation}
    dV=\Omega r^2 dr
    \label{eqvolume}
\end{equation}
where $r$ is the distance from the Sun.
 Therefore we define this number of stars as
\begin{equation}
    N(<r_{max}) = \int_0^{r_{max}} \rho(r,l,b) \; \; dV
    \label{eqnrmax}
\end{equation}
where $r_{max}$ is the maximum distance from the Sun that is defined from the observation depth.

\subsection{Density law of YSOs in the Galactic Thin Disk}
In the Galactic thin disk, the star density distribution or number density, 
usually called the density profile or density law, 
can be described by a double exponential \citep[e.g.][]{cabr05}, as follows:
\begin{equation}
    \rho(r,l,b)=A \times exp \left(-\frac{r |sin b|}{h} - \frac{R}{r_1}\right) 
    \label{eqrho}
\end{equation}
where  $A$ is the total density normalization (to be computed below), $h$ is the thin disk scale height that we assume equal to 300\,pc \citep{blan16},
$R$ is the Galactocentric distance, and 
$r_1$ is the thin-disk radial scale length that we assume equal to 2.6\,kpc
\citep{freu98}. The Galactocentric distance $R$ can be expressed in terms of $r$, $l$, and Galactic Center distance from the Sun $D$, as:

\begin{equation}
    R = \sqrt{(D - r \cos l)^2 + (r \sin l)^2}.
\end{equation}

To derive the density normalization constant $A$, we 
integrate the thin-disk density $\rho$ over the whole Galaxy, to obtain the total number of stars as:
\begin{equation}
\label{equn}
     N = 4 \pi A h r_1^2
\end{equation}


Recent results by \citet[][see their Table 2]{giam21} show that, within a factor of two, we can assume in the Galaxy a constant star formation rate. Therefore, 
in the whole Galaxy the total number of stars with ages t$<$10\,Myr, N$_{\rm yng}$, can 
be estimated as 
\begin{equation}
    {\rm N}_{\rm yng}=\frac{t_{\rm yng}}{t_{\rm MW}} \times {\rm N}_{\rm tot}
\end{equation}
where 
$t_{\rm yng}$=10\,Myr is the upper age limit for the youngest stellar component considered here,
$t_{\rm MW}$ is the Galaxy age that we assume to be $\sim$10\,Gyr and
N$_{\rm tot}$ is the total number of the stars in the Galaxy roughly equal to 10$^{11}$. With these assumptions N$_{\rm yng}$ amounts to 10$^8$ stars. Moreover, to take into account that we are going to miss about 20\% of the IMF below 0.3 $M_{\odot}$  \citep{weid10},
we should add a corrective factor of 0.8 for the number of actually observable stars.

Therefore, we may reasonably approximate in the Equ.\,\ref{equn} for the total thin-disk content $N \sim 0.8 \times N_{yng}$, from which $A \sim \frac{0.8 \times 10^8}{4 \pi h r_1^2}$, to be used in Eq.\,\ref{eqrho}. 

\subsection{Rubin LSST accessible volume}
The element of volume within a given solid angle $\Omega$,
as well as the density law $\rho$, defined by 
the  equations\,\ref{eqvolume} and \ref{eqrho}, respectively, 
depend on the distance from the Sun. 
At a given mass and age and for a given extinction,
the maximum  distance that can be achieved with Rubin LSST data  is a function of the photometric observational depth and thus of the adopted
observing strategy.
To evaluate such distance, and thus to define such volume, our metrics has been defined with the following
criteria:
\begin{itemize}
    \item use of the Rubin LSST gri filters, assuming to use 
    the $r-i$ vs. $g-r$ diagram to photometrically   select the YSOs,
    as in \citet{dami18,pris18,venu19}. 
    This is a restrictive assumption, since more than 3 Rubin LSST filters
    will be used for the selection;
    \item use of a dust map  to take into account the non-uniform
    extinction pattern that characterizes the GP, in particular towards the
    Galactic Center;
    \item estimate the maximum distance that will be achieved 
    assuming different \texttt{OpSims}. 
\end{itemize}



The definition of the metrics is therefore based on the following rationale:
specify the desired accuracy magnitude/$\sigma$ = 5 for gri filters
and compute the corresponding limiting magnitudes from coadded images; correct for extinction the apparent magnitudes using a dust map; compute the maximum distance at which a 10 Myr old
star of 0.3 M$_\odot$ can be detected, assuming the absolute magnitudes in the gri filters, M$_g$=10.32, M$_r$=9.28 and M$_i$=7.97,
predicted for such a star by a 10 Myr solar metallicity isochrone \citep{togn18};
calculate the corresponding volume element within a Nside=64 Healpix;
integrate the young star density within the volume element, through 
 equation\,\ref{eqnrmax}.

\subsubsection{Dust map}
The dust extinction is an increasing function of the 
distance and then nearby stars are affected by E(B-V) values smaller than those of more distant stars.
The \citet{schl98}  map has a massive use in Astronomy, but  provide an integrated extinction along a line of sight at the maximum distance
corresponding to the adopted 100\,$\mu$m observations.
As a consequence, they are more useful for extragalactic studies \citep{amor21}.
Assuming a 2D dust extinction  
i.e. a map that depends only on the positions (e.g. the galactic coordinates $l, b$) leads to a significant overestimation of the real 
extinction at different distances, especially in the directions
with A$_{\rm V}>$0.5 \citep[e.g][]{arce99,amor05}, such as in the GP
and, in particularly, in the Galactic spiral arms. 

In the last years, many efforts have been done to derive 3D extinction 
maps that take into account the strong dependence on the distance, as summarised in \citet{amor21}.

\begin{figure*}[htp]
 \includegraphics[scale=0.3]{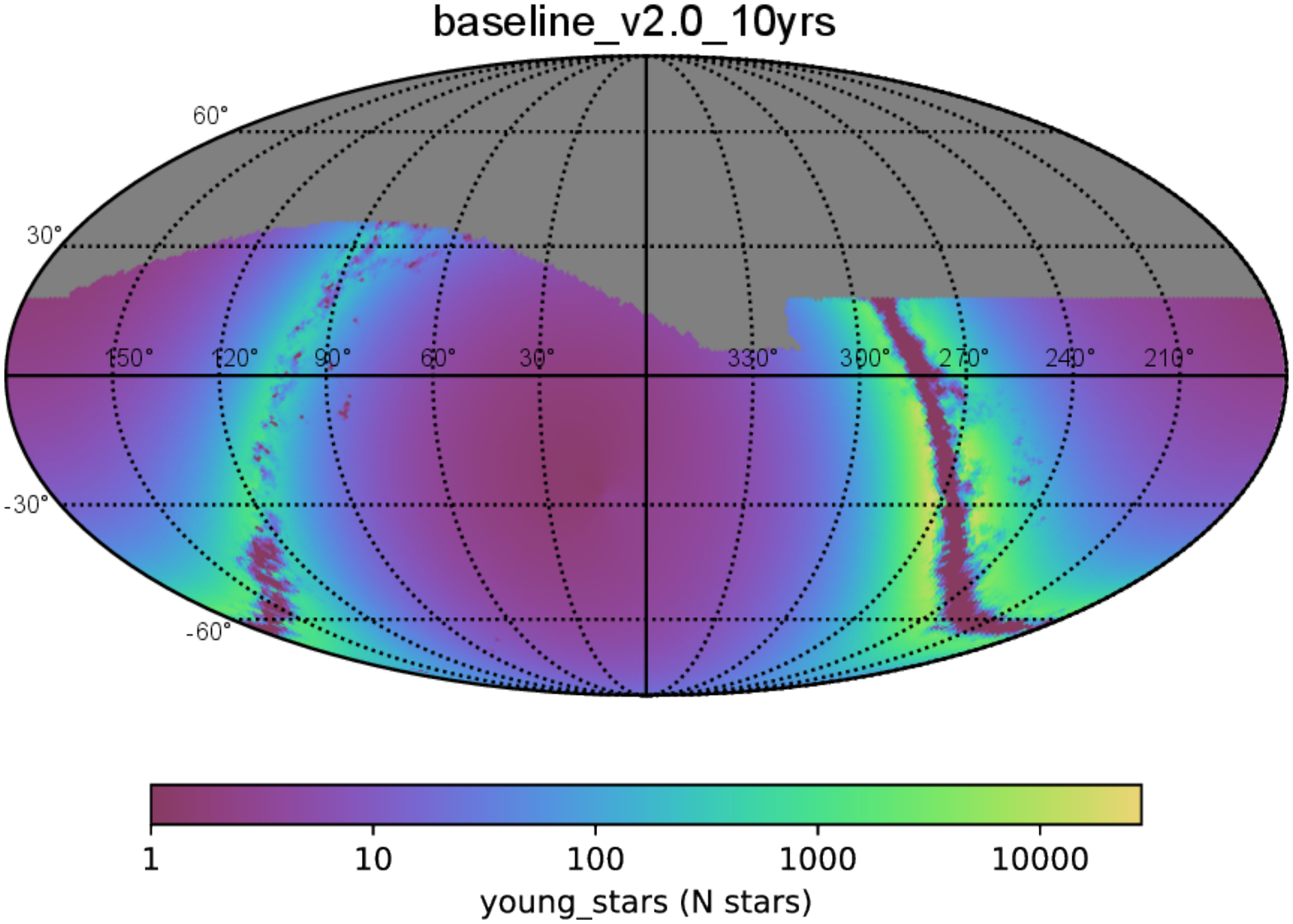}
 \includegraphics[scale=0.3]{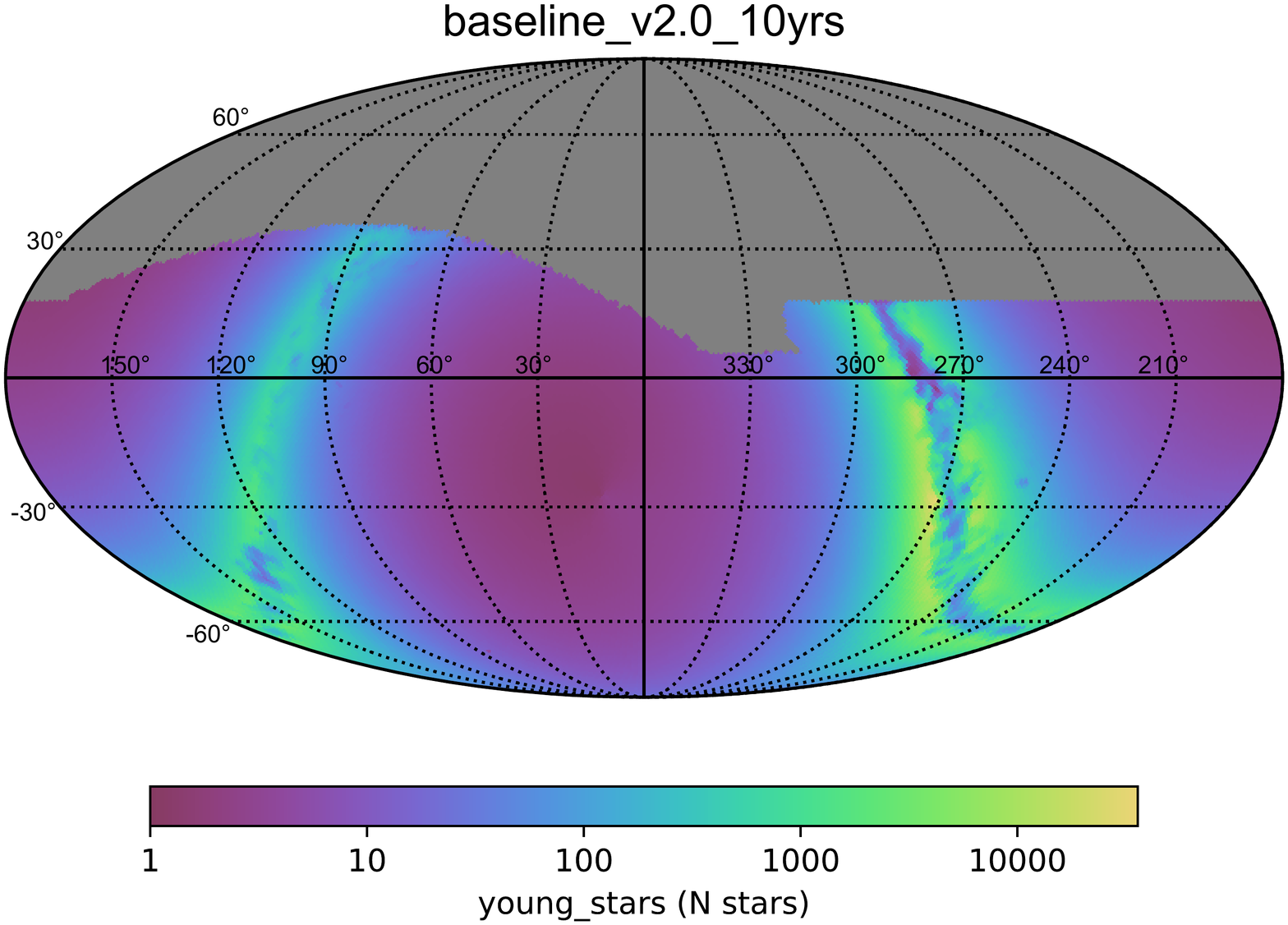}
 \caption{Map of the number of detectable YSOs per HEALPix computed  with our metrics, using the 
  2D dust map (left panel) and the 3D dust map (right panel). In both
 cases the  \texttt{baseline\_v2.0\_10yrs} has been adopted.
 The sky is shown in the equal-area Mollweide projection in equatorial coordinates and the HEALPIX grid  resolution is $N_{side}=64$.}
\label{fig:dustmapcomp}
\end{figure*}

 A 3D dust map,  appropriated 
for the Rubin LSST metrics,
 has been developed (Mazzi et al. in preparation) by using the \citet{lall19} map, the 
\texttt{mwdust.Combined19} map, based on the \citet{drim03,gree19,mars06} maps, combined with the method described in \citet{bovy16},
 and the  \citet{plan16} map Planck.
The 3D map has been integrated
in the MAF  as \texttt{DustMap3D} to distinguish  it from the default Rubin 2D \texttt{Dust\_values} based on \citet{schl98} map.
The two maps can be imported into MAF with the following commands: 
\begin{verbatim}
from rubin_sim.photUtils import Dust_values
from rubin_sim.maf.maps import DustMap3D.
\end{verbatim} 

To test the effect on our metrics of using the 2D or the 3D dust map and select the one predicting the more realistic results, we computed our metrics by considering the default 2D dust map implemented within the MAF, as well as the new 3D dust map.
Fig.\,\ref{fig:dustmapcomp} (left panel) shows the results of our metrics  on the OpSim simulation named \texttt{baseline\_v2.0\_10yrs},
obtained by assuming the 2D  dust map, while Fig.\,\ref{fig:dustmapcomp} (right panel) shows the corresponding result, assuming the 3D dust map.

The results point out that by assuming the 3D distant-dependent extinction map,  
the number of detectable YSOs 
decreases only smoothly towards the inner GP. On the contrary, the sharp decrease in the detected YSOs in the inner GP  is evidence of an unrealistic star distribution due to the adoption of an unsuitable overestimated extinction map (2D dust map). Therefore, 
 the final version of the code defining our metrics includes the  
  3D dust map.

%
\subsubsection{Crowding effects}
 Our metrics has been defined by assuming  the formal limiting magnitude at 5$\sigma$ photometric precision, but in several wide
areas of the Galactic Plane, the limiting magnitude should be set by the photometric errors due to the crowding.
To take into account such effect a specific  crowding metrics has been developed within MAF based on
 the TRILEGAL stellar density maps \citep{dalt22} to compute the errors 
 that will result from stellar 
 crowding\footnote{see \url{https://github.com/LSST-nonproject/sims_maf_contrib/blob/master/science/static/Crowding_tri.ipynb}}.
 
 Such crowding metrics has been included in our code  since we expect that the observations become severely incomplete when the photometric errors due to the crowding exceed 0.25 mag. These 
 photometric errors are derived using the formalism by \citet{olse03} and the 0.25 mag criterion has been empirically checked using deep observations of the Bulge (Clarkson et al., in preparation for this ApJS special issue). 
 Therefore, to detect the  faintest stars in 
 very crowded regions we used the minimum (brightest) magnitude between those obtained with \texttt{maf.Coaddm5Metric()} and \texttt{maf.CrowdingM5Metric(crowding\_error=0.25)}.
 
\begin{figure}[htp]
 \includegraphics[scale=0.3]{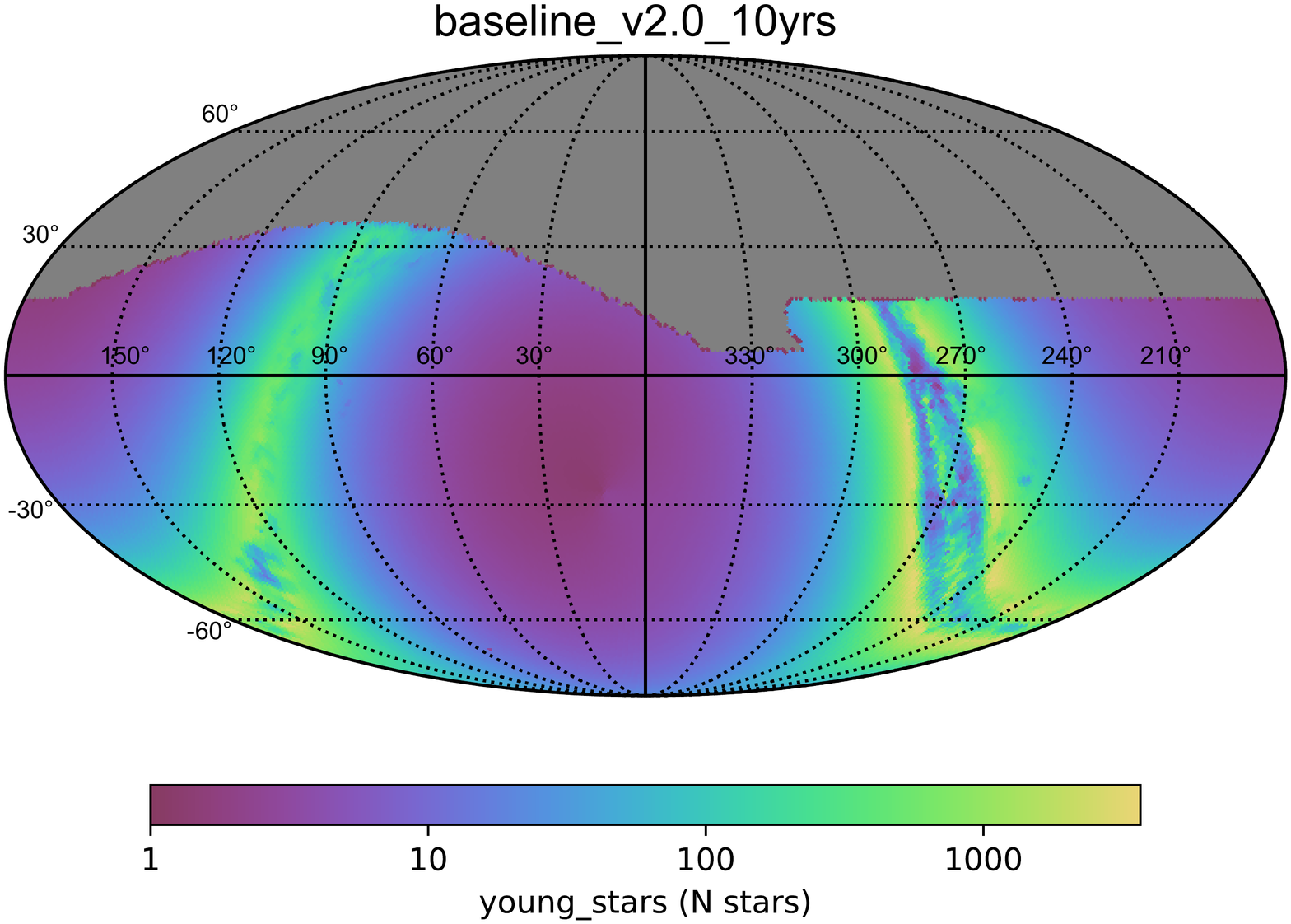}
\caption{
 Map of the number of detectable YSOs per HEALPix computed  with our metrics, using the \texttt{baseline\_v2.0\_10yrs},  the 3D dust map and the crowding metrics.  The sky is shown in the equal-area Mollweide projection in equatorial coordinates and the HEALPIX grid  resolution is $N_{side}=64$.}
\label{fig:dustmapcompcrow}
\end{figure}

 The limiting magnitudes achieved after including the crowding effects can be $\sim$2-3 magnitudes brighter than those obtained without
 considering confusion effects. As a consequence,  the number of YSOs detected by including the confusion metrics, shown in Fig.\,\ref{fig:dustmapcompcrow},  can be significantly smaller 
 than that predicted by the metrics that does not include such effect, shown in 
 Fig\,\ref{fig:dustmapcomp}, right panel.

 We note that the two-pronged fork feature in the map shown in Fig.\,\ref{fig:dustmapcompcrow}
 is due to the  fact that in the layers immediately above and below the GP, the crowding is the dominant effect
and the number of possible detections goes down here from more than 10000 to surprisingly low values of 10-100 (sources/HEALPix).
On the contrary,  
in the central layer of the GP,  the 
extinction is the dominant effect, while the crowding 
effect is negligible since  a smaller number of stars is visible. As a consequence, after applying the crowding metrics, the
 number of detections  in this central layer  remains almost constant 
 (around 1000 sources)\footnote{note the different color scales
 in the maps}, while   the missed detection fraction is significantly larger in the  layers immediately higher and lower than the GP. 
 This result is consistent with what recently pointed out by \citet{cant23} for the Gaia data.

\section{Results}
\begin{figure*}[htp]
\gridline{\fig{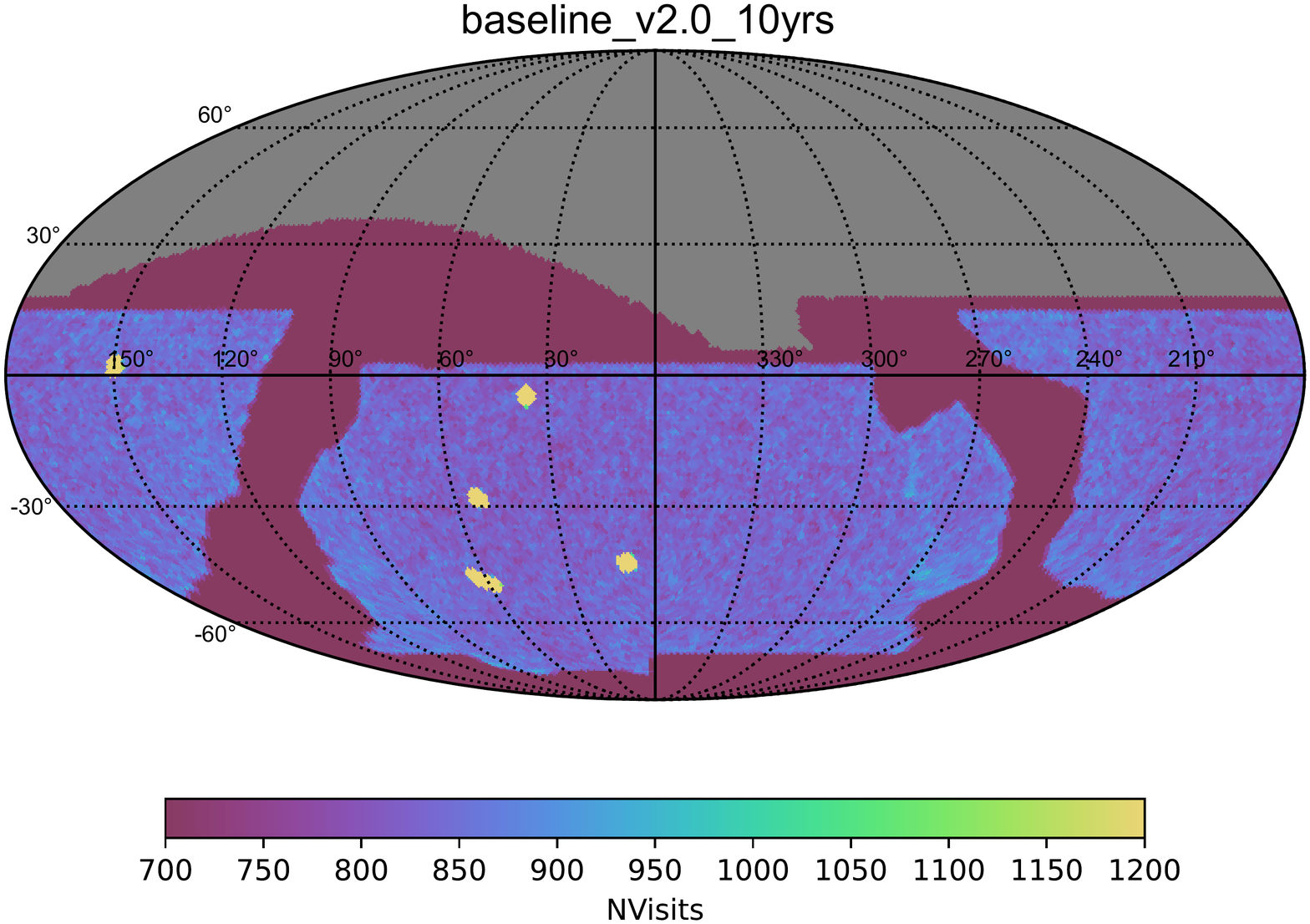}{0.45\textwidth}{(a)}
          \fig{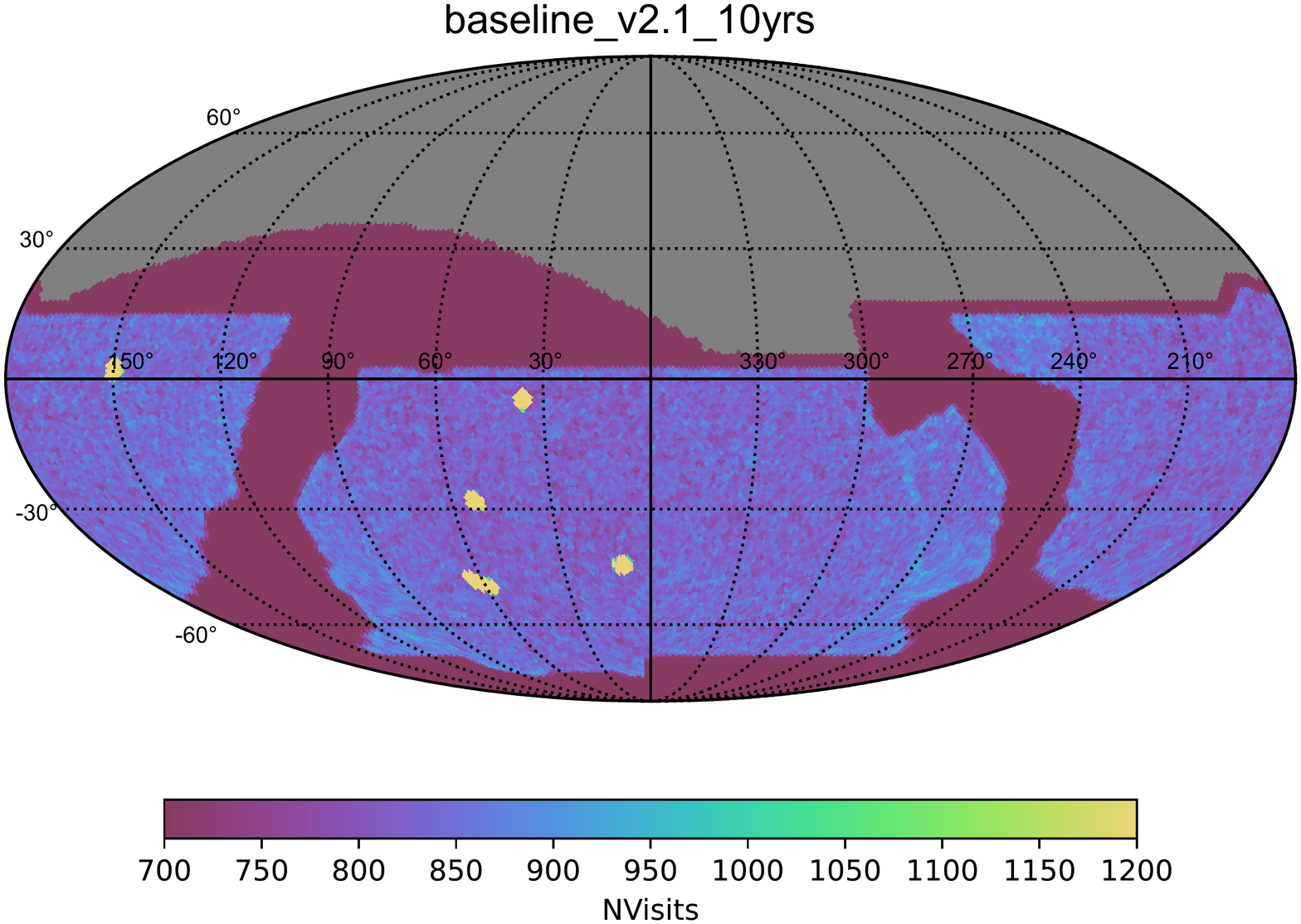}{0.45\textwidth}{(b)}
          }
\gridline{\fig{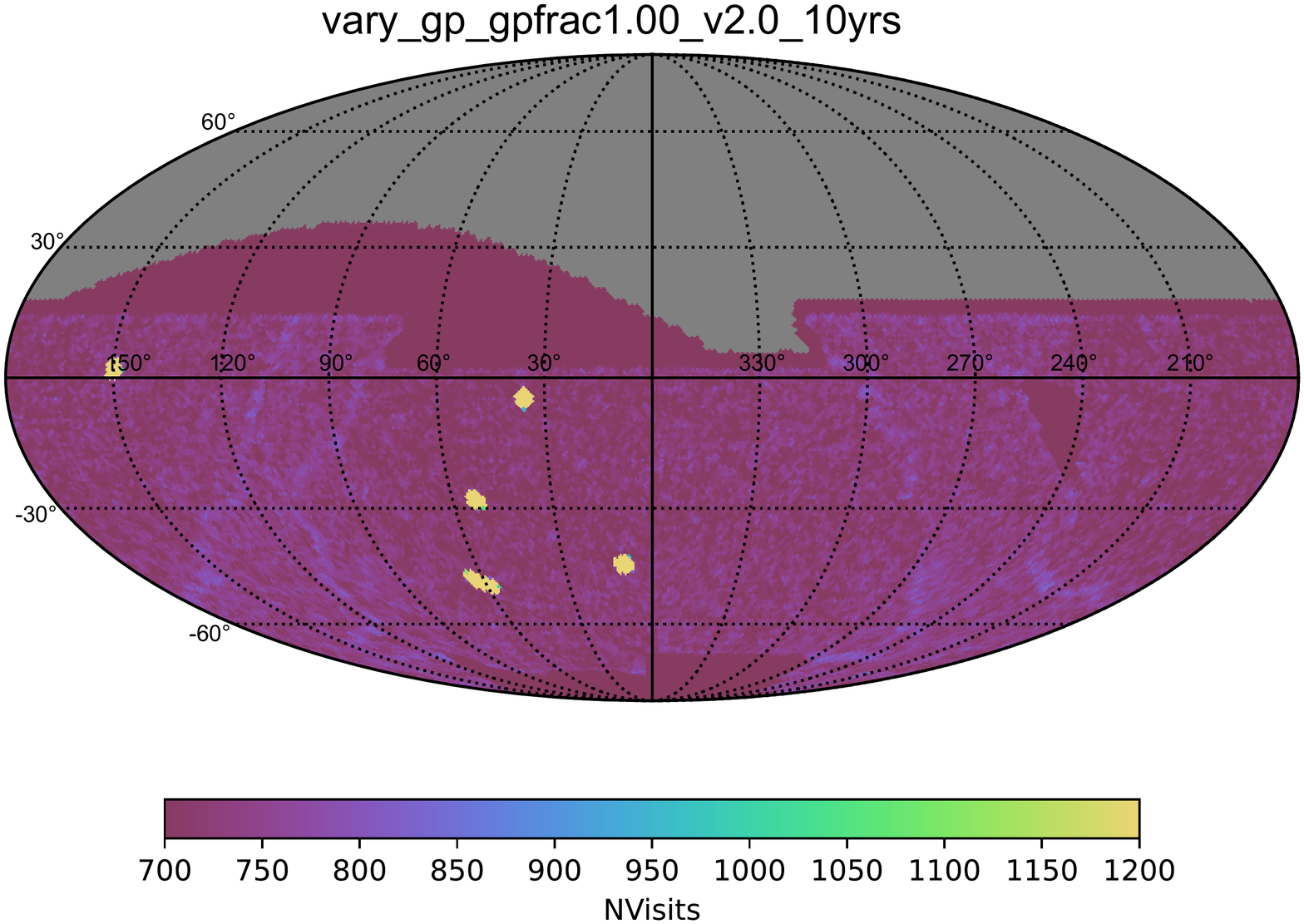}{0.45\textwidth}{(c)}
          \fig{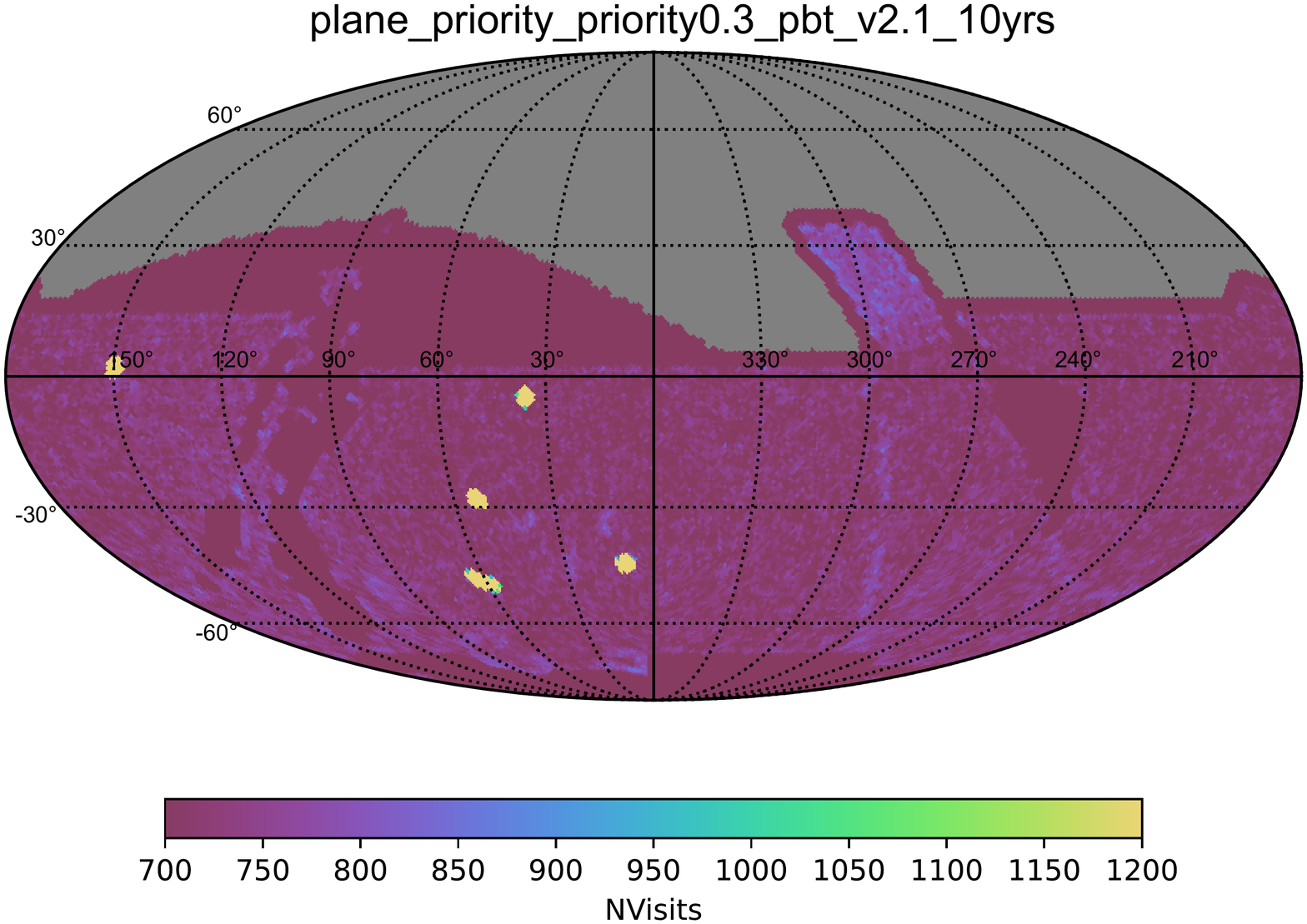}{0.45\textwidth}{(d)}
          }
\caption{Map of the number of visits per HEALPix planned for the four representative \texttt{OpSim} 
relevant for our science case. The small yellow patches are the Deep Drilling Fields (DDFs) to which an higher number of observations is assigned. Note as the Rubin-LSST footprint of the baselines (panels a and b) is significantly different with respect to that of the \texttt{vary\_gp\_gpfrac1.00\_v2.0\_10yrs}
and the \texttt{plane\_priority\_priority0.3\_pbt\_v2.1\_10yrs OpSim} (panels c and d),  where  the number of visits at the WFD level is assigned
also to the high dust extinction GP area.
The sky is shown in the equal-area Mollweide projection in equatorial coordinates and the HEALPIX grid  resolution is $N_{side}=64$.
}
\label{fig:nvisits}
\end{figure*}
The Python code that computes the metrics described in the previous sections is publicly available in the central
   \texttt{rubin\_sim}  MAF metrics  
repository\footnote{\url{https://github.com/lsst/rubin_sim/blob/main/rubin_sim/maf/maf_contrib/young_stellar_objects_metric.py}}
   
   with the name 
   \texttt{YoungStellarObjectsMetric.py}.

To evaluate the impact of the different family sets of \texttt{OpSim} on our science case, we considered the current state of LSST 
v2.0 and v2.1 \texttt{OpSim} databases and metrics results\footnote{available at 
\url{http://astro-lsst-01.astro.washington.edu:8080}}. In particular, we considered the
following   runs:
\begin{itemize}
    \item \texttt{{\bf baseline\_v2.0\_10yrs}},  the reference benchmark survey strategy in which the WFD survey footprint has been extended to include the central GP and bulge, and is defined by low extinction regions. Such configuration  includes also five Deep Drilling Fields and additional mini-survey areas of the North Ecliptic Plane, of the GP, and of the South Celestial Pole.
    \item \texttt{{\bf baseline\_v2.1\_10yrs}}, in which the Virgo cluster and the requirement on the     seeing fwhmEff $<$ 0.8\sec\, for the images in r and i bands have been added with respect to the \texttt{baseline\_v2.0\_10yrs}. 
    \item {\bf Vary GP} family of simulations in which the fraction of the amount of survey time spent on covering the background (non-WFD-level) GP area ranges 
    from 0.01 to 1.0 (labeled as frac0.01 - frac1.00), where 1.0 corresponds to extending the WFD cadence to the entire GP. The baseline characteristics, including the ratio of visits over the remainder of the footprint, are kept the same. In our case,
    we considered the \texttt{vary\_gp\_gpfrac1.00\_v2.0\_10yrs} survey, planning the maximum amount of time on the GP.
    \item {\bf Plane Priority}, the family of \texttt{OpSim}  that uses the GP priority map, contributed by LSST SMWLV \& TVS science collaborations, as the basis for further variations on GP coverage.
    For this family of simulations, different levels of priority of the GP map are covered at WFD level, 
    unlike the baseline and Vary GP family where only 
    the "bulge area" is covered at WFD level. 
     In our case, we  considered the 
    \texttt{plane\_priority\_priority0.3\_pbt\_v2.1\_10yrs} survey,
    since it  maximises the results of our metrics. 
\end{itemize}

\begin{figure*}[!htp]
 \gridline{\fig{dbFiles_baseline_v2_0_10yrs_young_stars_HEAL_SkyMap_p.pdf}{0.45\textwidth}{(a)}
    \fig{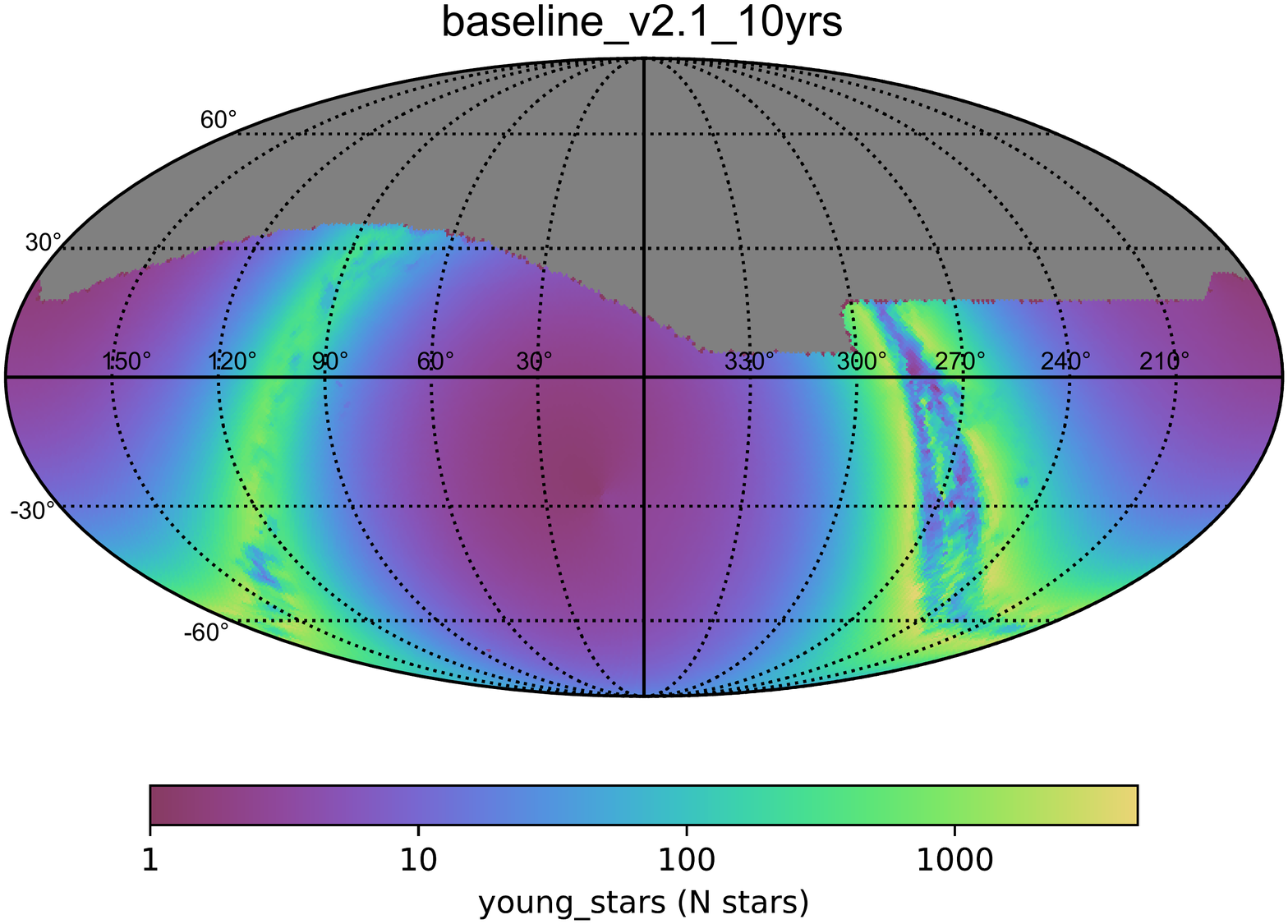}{0.45\textwidth}{(b)}
          }
\gridline{\fig{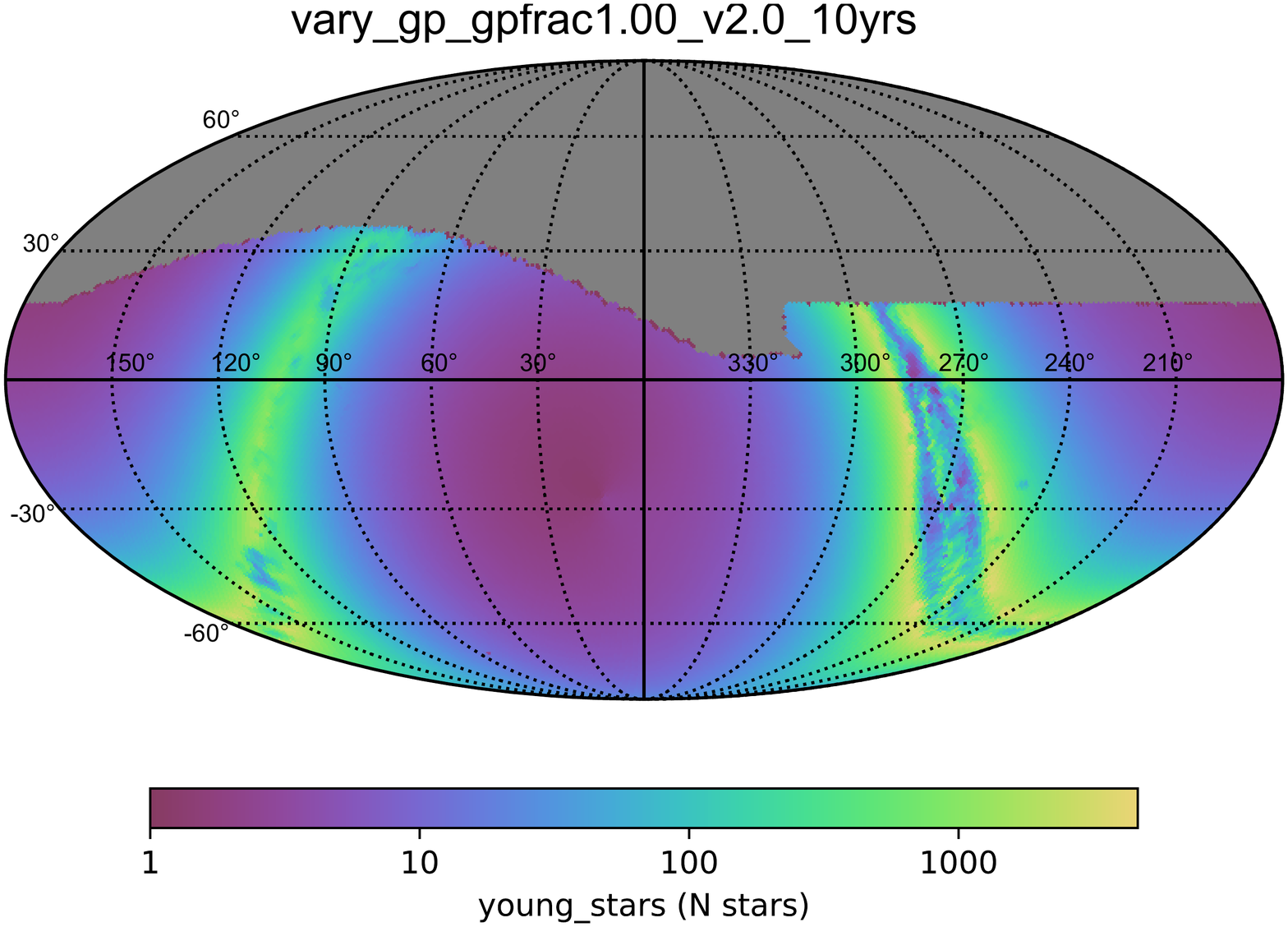}{0.45\textwidth}{(c)}
\fig{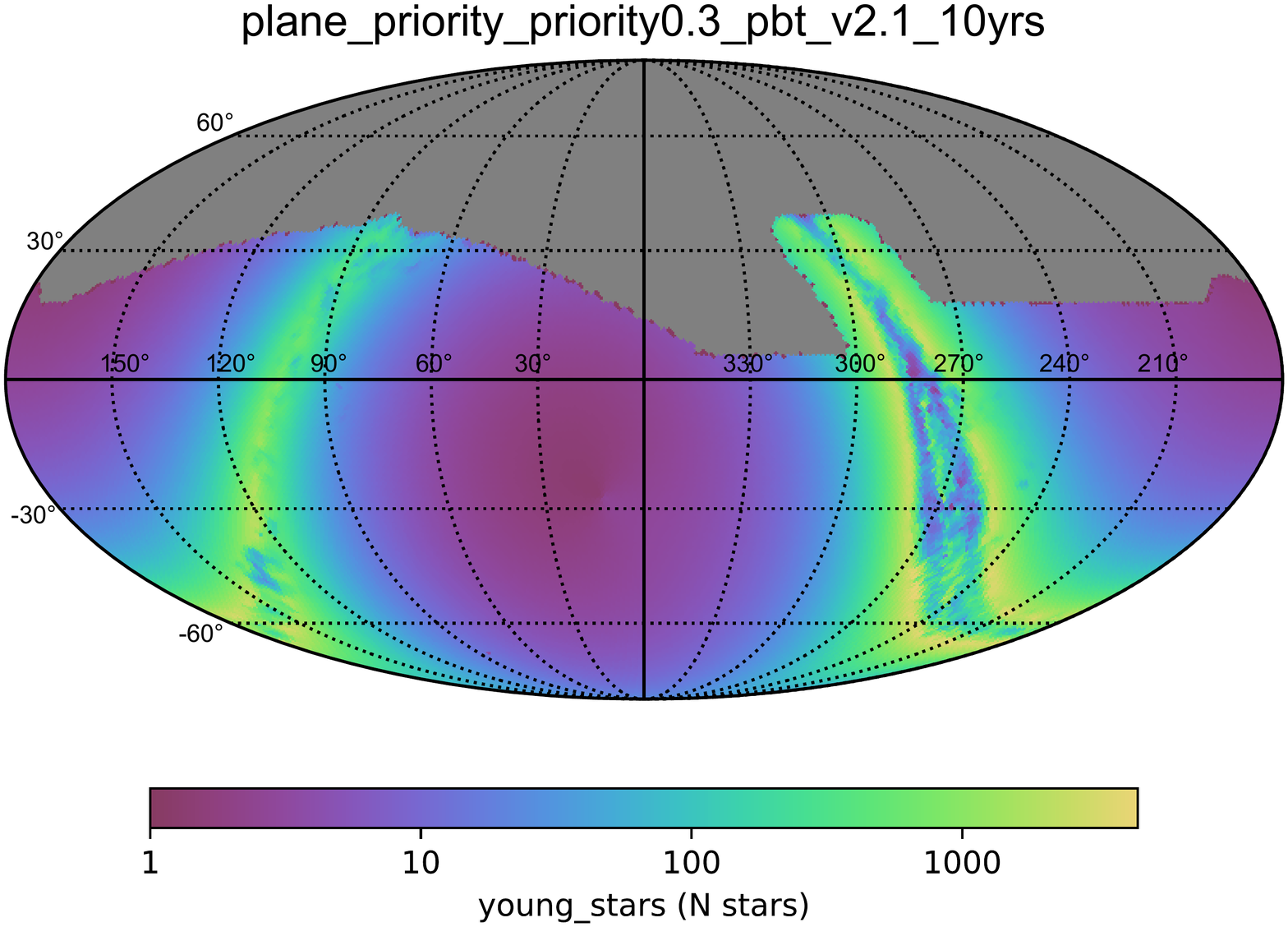}{0.45\textwidth}{(d)}
          }
\caption{Map of the number of YSOs per HEALPix as computed 
by our metrics, adopting the  \texttt{OpSim} surveys listed
in Table\,\ref{tab:resultsmaf} and indicated by the names superimposed on top of each panel. The sky is shown in the equal-area Mollweide projection in equatorial coordinates and the HEALPIX grid  resolution is $N_{side}=64$. Note that the map in panel (a) is identical to the map shown in Fig.\,\ref{fig:dustmapcompcrow}.}
\label{fig:resultsmafold}
\end{figure*}

To evaluate the best observing strategy to be adopted for the YSO science,
we defined as Figure of Merit (FoM), the ratio between 
the number of young stars detected with a 
given \texttt{OpSim} with respect to the number of young stars detected
with the \texttt{baseline\_2.0\_10yrs} survey, this latter taken
as reference.
The map of the number of visits planned for these four representative \texttt{OpSim} surveys are shown in Figure\,\ref{fig:nvisits}, while
the results of the metrics, i.e. the  number of YSOs with ages $t<10$\,Myr
and masses $>0.3$\,M$_\odot$, obtained  with these four representative \texttt{OpSim} surveys are
given in Table\,\ref{tab:resultsmaf} and shown in Figure\,\ref{fig:resultsmafold}. 
The FoM obtained with our metrics using the four \texttt{OpSim} described before are also given in 
Table\,\ref{tab:resultsmaf}.  Both, the number of YSOs and the FOM predicted by the metrics not including and including the crowding effects are
given in order to quantify the differences due to the confusion effects.
\begin{table*}[]
    \centering
    \begin{tabular}{c c c c c}
   \hline 
    \hline 
\texttt{OpSim} ID &  N   &  FoM & N$_{\rm Crow}$ &  FoM$_{\rm Crow}$ \\
       \hline 
\texttt{baseline\_v2.0\_10yrs}    &                      
8.08$\times 10^6$ & 1.00 & 4.84$\times 10^6$ & 1.00 \\  
\texttt{baseline\_v2.1\_10yrs}                          & 
8.10$\times 10^6$ & 1.00 & 4.87$\times 10^6$ & 1.01\\ 
\texttt{vary\_gp\_gpfrac1.00\_v2.0\_10yrs}              & 
8.92$\times 10^6$  & 1.10 & 5.58$\times 10^6$ & 1.15\\ 
\texttt{plane\_priority\_priority0.3\_pbt\_v2.1\_10yrs} & 
9.51$\times 10^6$ & 1.18 & 6.02$\times 10^6$  & 1.24 \\ 
    \hline
    \end{tabular}
\caption{Number of YSOs with ages t$<$10\,Myr and 
masses $>$0.3\,M$_\odot$ that can be detected at distances $<r_{max}$
set by different \texttt{OpSim} surveys,
and relative FoM values, estimated  by neglecting the crowding
metrics (N and FoM columns) and by including it
(N$_{\rm Crow}$ and FoM$_{\rm Crow}$ columns).}
    \label{tab:resultsmaf}
\end{table*}

\section{Discussion and conclusions}
In this paper we presented a metrics aimed to estimate the number of
YSOs we can discover with Rubin LSST static science data. 
To take into account the effects due to the extinction, we evaluated the
metrics by using both, the default LSST MAF 2D  dust map  and the 3D dust map
where a more realistic dependence on the distance is included.
The results of this comparison show that the 2D dust map overestimates E(B-V) values for close (few kpc) accessible stars, even in the direction of the GP. 
For this reason, the final metrics has been implemented by considering  the 3D dust map integrated into MAF.

To evaluate the faintest magnitudes that can be attained in crowded regions, a specific crowding metrics has been
imported in the code. 
In order to quantify the crowding effects,  we evaluated our metrics also by neglecting the crowding effects.
The results  
show that,  if the  crowding metrics is included, the number of 
detected YSOs (N$_{\rm Crow}$ in Table\,\ref{tab:resultsmaf}) is a factor 
 0.6-0.63 smaller than the number of YSOs predicted if the crowding is 
 not considered (column N in Table\,\ref{tab:resultsmaf}). Since the most realistic predictions are those obtained by including the
 crowding effects, the final version of the code
 that defines the YSOs metrics includes
the crowding metrics.

The impact of changing from the baseline v2.0 to v2.1 is 
negligible, while
adopting the \texttt{vary\_gp\_gpfrac1.00\_v2.0\_10yrs} survey,
the only one available  covering 100\% the GP with WFD exposure times, 
the predicted number of YSOs (t$<$10\,Myr) down to 0.3\,M$_\odot$
is a  factor 1.15 larger than that predicted adopting the
\texttt{baseline\_v2.0\_10yrs}, leading to a gain of 0.74$\times 10^6$ new discovered YSOs at very large distances. 
An even higher impact can be obtained by adopting the most recently
contributed 
\texttt{plane\_priority\_priority0.3\_pbt\_v2.1\_10yrs} predicting
different levels of priority of the GP map. 
With this survey, the predicted number of YSOs is a factor 1.24 higher
with respect to the baseline, corresponding to a gain of  
$\sim 1.17 \times 10^6$ additional YSOs.

\begin{figure}[htp]
 \includegraphics[scale=0.6]{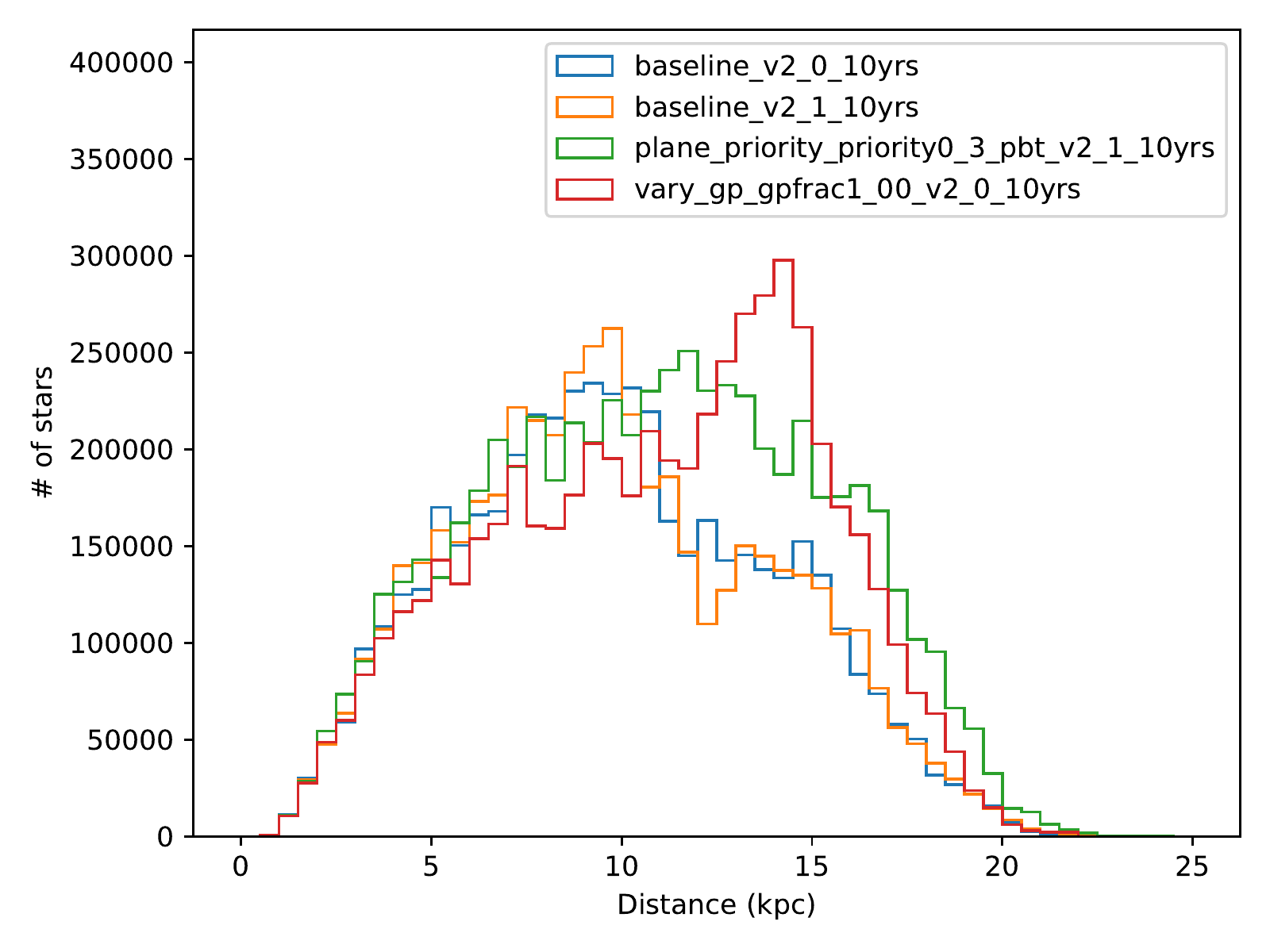}
 \caption{Number of detected YSOs  as a function of the distance for the four \texttt{OpSim} indicated in Table\,\ref{tab:resultsmaf}.}
\label{fig:histdist}
\end{figure}

To determine at what distance and beyond which distance  from the Sun this gain is
most significant, we included in our metrics also the  computation 
of the maximum distance that can be reached for a 10\,Myr old star of 0.3\,M$_\odot$
within each Healpix.
We derived the distribution of such distances, weighted by the 
number of YSOs that can be detected in each line of sight.
The distributions obtained for  the four \texttt{OpSim} considered in this work
are shown in Fig.\,\ref{fig:histdist}. 

We note that 
the number of YSOs detected at distances from the Sun larger than $\sim$10\,kpc 
is significantly larger for the \texttt{OpSim}  \texttt{plane\_priority\_priority0.3\_pbt\_v2.1\_10yrs} 
and  \texttt{ vary\_gp\_gpfrac1.00\_v2.0\_10yrs} 
than for the two baselines.  The curve of the \texttt{ vary\_gp\_gpfrac1.00\_v2.0\_10yrs} 
\texttt{OpSim} sits below the baseline curves for d=7-10 kpc. This could be due 
to the fact that in the two baselines, the number of visits in the regions outside the
GP is higher (800-900) with respect to that planned for the \texttt{ vary\_gp\_gpfrac1.00\_v2.0\_10yrs}  
(700-800), where the coverage is uniform.  
This comparison clearly shows that the science enabled by adopting an
observing strategy  consistent with the 
\texttt{plane\_priority\_priority0.3\_pbt\_v2.1\_10yrs} 
or  \texttt{ vary\_gp\_gpfrac1.00\_v2.0\_10yrs}  \texttt{OpSims} 
would be precluded if we followed the prescriptions of other simulations.
These  results highlight the greatest impact of Rubin LSST WFD cadence data
extended to the GP, mainly arising from the unknown farthest young stars that will be discovered.
We note that the two baselines and the  \texttt{ vary\_gp\_gpfrac1.00\_v2.0\_10yrs} \texttt{OpSims}
show an asymmetric distribution with a peak that is around 9\,kpc and 14\,kpc, respectively. 
The  \texttt{plane\_priority\_priority0.3\_pbt\_v2.1\_10yrs} \texttt{OpSim}
shows a more smoothed and symmetric shape, that is likely due to the 
different levels of priority given to the GP for this survey strategy.

We retain that a gain of 24\% in the number of discovered YSOs, mainly
arising at distances larger than 10\,kpc, 
represents a strong justification to extend the WFD strategy to the GP, 
and to ensure the uniformity of the observing strategy in the Galaxy. 

Finally, we note that our metrics scales more strongly with the area 
rather than with the number of visits per pointing and therefore, in case the
\texttt{plane\_priority\_priority0.3\_pbt\_v2.1\_10yrs} survey
cannot be adopted, we advocate the adoption of the 
\texttt{ vary\_gp\_gpfrac1.00\_v2.0\_10yrs},  to preserve the uniform coverage of the GP.

We stress that our request is to extend the number of visits planned for the
WFD  to the GP 
but without constraints on the temporal observing cadence, since our analysis is based on the 
use of the co-added gri Rubin LSST images.
However, different visits per night should be in different filters from the
  gri set.  The same applies in case 2 or 3 visits/night are made, as this would enhance the number
  but also the accuracy of color measurements available, and hence heightens the fraction of YSOs 
  that can be detected and characterized thanks to their color properties.
  The color accuracy is improved if the observations in different filters are taken close together in time and therefore 
  least affected  by star variability.

 The metrics described in this work has been developed assuming to use a photometric technique to statistically identify YSOs
mainly expected within SFRs as already done in \citet{dami18,pris18,venu19}. However, YSOs can also form  dispersed populations, mingled with the general thin disk population 
\citep[e.g.][]{bric19,luhm22}, and their identification can be hampered by the strong field star contamination. Such issue can be
{\bf overcome}  by also exploiting Rubin LSST accurate proper motion
and parallax measurements \citep{lsst17,pris18a}, as well as the most deep photometric near-IR photometric surveys, such as, for example, the 
VISTA Variables in the Via Lactea \citep{minn10}. Nevertheless,
 a spectroscopic follow up will be needed for 
a full characterization of the YSOs detected with Rubin LSST, for
which spectroscopic data can be achieved.
Forthcoming  spectroscopic facilities like WEAVE \citep{dalt20} and 
4MOST \citep{dejo22} will be crucial to this aim, at least within the
limiting magnitudes they can achieve.
  
The exquisite depth of the Rubin LSST coadded images represents a unique opportunity to study large-scale young structures and
to investigate the low mass stellar populations in/around the thin disk GP, where star formation mainly occurs. Our results suggest that 
in order to maximise the volume of detected YSOs  and the uniformity of
coverage of the large-scale young structures,
the amount of survey time dedicated to the GP should be equal to or comparable to the WFD level. These observations will allow us 
to better characterize the  Galactic structure and in particular the Norma, Scutum, Sagittarius, Local, Perseus and the Outer spiral arms, reported in \citet{reid14}.

\begin{acknowledgments}
This work was supported by the Preparing for Astrophysics with LSST Program, funded by the Heising Simons Foundation through grant 2021-2975, and administered by Las Cumbres Observatory.
AM acknowledges financial support from Padova University, Department of Physics and Astronomy Research Project 2021
(PRD 2021). L.V. is supported by the National Aeronautics and Space Administration (NASA) under grant No. 80NSSC21K0633 issued through the NNH20ZDA001N Astrophysics Data Analysis Program (ADAP).
\end{acknowledgments}
\bibliography{bibdesk}{}
\bibliographystyle{aasjournal}

\end{document}